\documentclass[twocolumn, times, tighten,twocolappendix]{aastex631}
\usepackage{graphicx,multirow,hyperref,url,color,xspace}
\usepackage{amsmath,amssymb}

\newcommand{\msun}{{\rm M}_{\sun}}

\newcommand{\g}{$\gamma$}

\newbox\grsign \setbox\grsign=\hbox{$>$} \newdimen\grdimen \grdimen=\ht\grsign
\newbox\simpropbox
\setbox\simpropbox=\hbox{\raise.5ex\hbox{$\propto$}\llap
   {\lower.5ex\hbox{$\sim$}}}\ht2=\grdimen\dp2=0pt

\defcitealias{Chiang02}{C02}
\newcommand{\chiang}{{\citetalias{Chiang02}}\xspace}
\defcitealias{Done12}{D12}
\newcommand{\done}{{\citetalias{Done12}}\xspace}

\begin{document}

\title{Corrections to Estimated Accretion Disk Size due to Color Correction, Disk Truncation and Disk Wind}
\shorttitle{Accretion Disk Size}

\author[0000-0002-0333-2452]{Andrzej A. Zdziarski}
\affiliation{Nicolaus Copernicus Astronomical Center, Polish Academy of Sciences, Bartycka 18, PL-00-716 Warszawa, Poland; \href{mailto:aaz@camk.edu.pl}{aaz@camk.edu.pl}}
\author[0000-0002-8231-063X]{Bei You}
\affiliation{School of Physics and Technology, Wuhan University, Wuhan 430072, China}
\author[0000-0001-7606-5925]{Micha{\l} Szanecki}
\affiliation{Faculty of Physics and Applied Informatics, {\L}{\'o}d{\'z} University, Pomorska 149/153, PL-90-236 {\L}{\'o}d{\'z}, Poland}

\begin{abstract}
We consider three corrections to the disk sizes estimated at a given frequency using accretion models. They are due to a color correction, a disk truncation at an inner radius larger than the innermost stable circular orbit, and disk winds, which we apply to the standard disk model. We apply our results to the estimates of the disk sizes based on microlensing. We find these three effects combined can explain the long-standing problem of the disk sizes from microlensing being larger than those estimated using the standard disk model (i.e., that without accounting for the above effects). In particular, an increase of the color correction with the increasing temperature can lead to a strong increase of the half-light radius even if this correction is close to unity at the temperature corresponding to an observed frequency. Our proposed formalism for calculating the half-light radius also resolves the long-standing issue of discrepancies between the disk size estimates based on the accretion rate and on the observed flux.
\end{abstract}

\section{Introduction}
\label{intro}

About a hundred of distant quasars have so far been observed to have multiple images due to gravitational lensing by a galaxy in the line of sight. A famous example is the Einstein cross source, G2237+0305, which has four gravitationally lensed images \citep{Schneider88}. Then, stars in the lensing galaxies cause microlensing \citep{Paczynski86} of each of the image, which effect allows us to determine the size of the source of light, in particular, an accretion disk surrounding the quasar black hole (BH), see \citet{Wambsganss06} for a review.

A number of papers have reported that the sizes estimated by O/IR monitoring observations are larger by a factor of a few than those predicted by the standard accretion disk theory \citep{SS73} given the observed flux (e.g., \citealt{Pooley07,Dai10,Morgan10, Chartas16, Cornachione20}). Equivalently, the optical flux measured from the source is significantly lower than that from a standard geometrically-thin disk emitting at the wavelength of the observation. We consider here ways to explain this discrepancy.

In this work, we consider local color corrections to the disk blackbody emission of $f_{\rm col}\geq 1$. At a given observed flux and frequency, the size of the emitting region is $\propto f_{\rm col}^2$, and $f_{\rm col}\sim 2$ would be sufficient. The study of \citet{Hubeny01} shows that $f_{\rm col}\approx 1$ in the optical wavelengths. However, this assumes that there is no dissipation in disk surface layers. If this assumption is relaxed and a moderate dissipation in the surface layer is allowed, the above discrepancy can be resolved. Indeed, there have been a number of papers considering accretion disks supported by magnetic pressure, which are substantially hotter than the standard ones \citep{Begelman07,Salvesen16a}, see also \citet{Begelman17}, \citet{Dexter19}, \citet{Mishra20}. The effect of scattering atmospheres on the disk size estimates was also considered by \citet{Hall18}. 

Then, even if the color correction is $\sim$1 at the measured frequencies, its increase at higher frequencies can lead to a substantial increase of the half-light radius (as defined by \citealt{Mortonson05}), for which a half of the radiation at a $\nu$ is emitted interior of it, and a half, exterior to it. 

A second effect is a possible truncation of the disk at some inner radius. This is likely, given the X-ray emission observed from those quasars. If the disk is truncated, the half-light radius will be increased further. Third, there is a possible decrease of the accretion rate with the decreasing distance to the BH due to wind mass loss (e.g., \citealt{You16a}), which could, in principle, reconcile the theoretical predictions with the observations \citep{Li19, Sun19}. Here, we study the effects of all three of these phenomena in the framework of the standard accretion disk model.

\section{Disc size}
\label{disk}

\subsection{Formulae}
\label{formulae}

We consider a geometrically-thin, optically-thick, accretion disk \citep{SS73}. We assume the disk locally, at a radius $R$, emits a blackbody spectrum diluted by a color correction, $f_{\rm col}\geq 1$, which can be a function of the effective temperature, $T_{\rm eff}$ (see, e.g., \citealt{Hubeny01, Davis06, Davis19}). We assume the disk is not covered by a corona. Then the observed flux at an observed frequency, $\nu$, is given by,
\begin{align}
&\nu F_\nu =   {4 \pi h\nu^4 (1+z)^4 (G M)^2 \cos i \over D_L^2 c^6}\times\label{nufnu}\\
&\quad\int^{r_{\rm out}}_{r_{\rm in}}\!\! { r {\rm d}r\over f_{\rm col}\left[T_{\rm eff}(r)\right]^4\left\{\exp\left[h \nu(1+z)/k T(r)\right]-1\right\}},
\nonumber\\
\label{tcol}
&T(r)=f_{\rm col}\left[T_{\rm eff}(r)\right]T_{\rm eff}(r),
\end{align}
where $r\equiv R/R_{\rm g}$, $R_{\rm g}=GM/c^2$ is the gravitational radius, $M$ is the BH mass, $z$ is the redshift, $D_L$ is the luminosity distance, $i$ is the inclination at which the disk is observed, $r_{\rm out}$ is the disk outer (dimensionless) radius, and $r_{\rm in}$ is the inner radius down to which the above optically disk extends. This radius may correspond to the innermost stable orbit (ISCO), but it may be substantially larger if the disk is truncated. In fact, the presence of X-ray emission from quasars requires that a part of the available gravitational energy is converted into heating of a relativistic gas, capable of X-ray emission. The energy conservation requires that the same part is subtracted from the disk dissipation. This can be achieved by either some of the dissipation taking place at the corona at the expense of the underlying disk \citep{SZ94,Done06, You16b}, or by truncation of the optically thick disk at $R_{\rm in}$, below which it is replaced by a hot flow (e.g., \citealt*{NY94, ZG04,DGK07}), or by a combination of both. Hereafter, we assume that the disk is truncated. While this is not a unique possibility, it is consistent with the measured sizes of the X-ray emitting region in microlensing sources being much lower than those emitting in the optical band \citep{Mosquera13}. We also note that $D_L=D_{\rm A}(1+z)^2$, where $D_{\rm A}$ is the angular diameter distance, which allows for a minor simplification of the above expression. 

In the Newtonian limit and assuming the zero-torque inner boundary condition (see \citealt{Paczynski00} for arguments for the validity of this condition), the effective temperature is\footnote{We note the transfer of the angular momentum is responsible for the appearance of the factor of 3 in equation (\ref{teff}) regardless of the boundary condition, cf.\ equations (4.30) and (5.14--5.15) of \citet{FKR02}. The locally released gravitational energy would give only the factor of 1 in that equation. In the case of zero-stress inner boundary condition, the factor of 3 also corresponds to $\int_{r_{\rm in}}^\infty r^{-2} {\rm d}r=1/r_{\rm in}$ vs.\ $\int_{r_{\rm in}}^\infty r^{-2} [1-(r_{\rm in}/r)]^{1/2}{\rm d}r=(1/(3 r_{\rm in})$, independent of the value of $r_{\rm in}$. We note that this factor of 3 should appear in the denominator of equation (12) of \citet{Zdziarski21a}.}
\begin{equation}\label{teff}
T_{\rm eff}(r) = \left[ 3  c^6\dot M(r)\over 8 \pi \sigma G^2 M^2 r^3 \right]^{1/4} \left[1 - \left(r_{\rm b}\over r \right)^{1/2} \right]^{1/4}, \\
\end{equation}
where $\dot M(r)$ is the disk mass flow rate at $r$, $r_{\rm b}$ is here the radius at which the zero-torque boundary condition is imposed, and $\sigma$ is the Stefan-Boltzmann constant. We note that the dissipation may continue in a hot flow below a disk truncation radius, $r_{\rm in}$, which thus could be $> r_{\rm b}$. Therefore, we allow for $r_{\rm in}\geq r_{\rm b}$ in the examples given below.

We note that most of the published studies of the microlensing disk sizes neglected the changes of the color correction with the disk radius and neglected the boundary term in Equation (\ref{teff}). The advantage of that approach is the simplicity of the resulting formulae. For completeness, we give them in Appendix \ref{simple}. In that approach, there are two inferred characteristic sizes, the so-called theory size and flux size.

Here we use instead the half-light radius, $R_{1/2,\nu}$ (as defined by \citealt{Mortonson05, Pooley07}), at which a half of the emission at $\nu$ is emitted interior of $R_{1/2,\nu}$. This is a nearly model-independent quantity, and thus preferred in comparison of models with observations \citep{Mortonson05}. Its form assuming a constant $f_{\rm col}$ and neglecting the zero-stress boundary condition is given in Equation (\ref{xhalf}).

In our approach below, we calculate the half-light radius using directly equations (\ref{nufnu}--\ref{teff}), and taking into account relevant physical effects. In particular, the emitting region can be close to the disk inner boundary, in which case the boundary term needs to be taken into account. Also, the color correction depends, in general, on the local effective temperature. Furthermore, winds can be launched from the accretion disk accretion disks via a variety of mechanisms: line-driven, thermal and magnetic. Details of those mechanisms remain still uncertain. 

In their studies of the effect of winds on the measured size of disks, \citet{Li19} and \citet{Sun19} assumed the decrease with the decreasing radius of the mass flow rate due to the wind for a non-rotating BH to be $\dot M(r)=\dot M_{\rm in}{(r/r_{\rm in})}^s$, where $r_{\rm in}=6$, $\dot M_{\rm in}$ is the accretion rate onto the BH, and $s>0$ is a parameter. \citet{Li19} fitted this model to the sizes obtained from microlensing, and found that the value of $s$ as high as $\approx$1.3 was required. If the above functional dependence applies up to the outer truncation radius, which in AGNs is typically $\sim\! 10^3$--$10^4 R_{\rm g}$ (e.g., \citealt{Goodman03}, and fig.\ 6 in \citealt{You12}), the mass flow rate supplied to the disk is $\sim\! 800$--$1.5\times 10^4$ higher than that accreting onto the BH, which we consider highly unlikely. In the case of sources emitting at a substantial fraction of the Eddington luminosity, e.g., $\sim$1/4--1 (typical for microlensing sources, \citealt{Kollmeier06, Davis11}), such an increase of $\dot M$ with the increasing radius would cause the accretion rate to become progressively more super-Eddington with the increasing radius, while there have no indications of such rates (see also \citealt{Davis11}). Also, the calculations of the emitted spectrum in such case assuming the thin disk become not self-consistent. In the case of accreting X-ray binaries, \citet{Coriat12} found that estimating the average mass transfer rate from the donor based on the observed X-ray luminosity assuming the standard accretion efficiency, i.e., neglecting the effect of winds in reducing the BH accretion rate, gives a very good agreement with the prediction of the disk instability model of the X-ray transients \citep{Dubus01}. Therefore, we believe it is highly unlikely that the presence of disk winds results in a radial dependence of the accretion rate as strong as the above power-law. Therefore, we consider alternative descriptions of the effect of winds on $\dot M(r)$. A number of them were listed by \citet{You16a}. Among those, we have chosen the most physically-motivated one, following from the jet/disk outflow model of \citet{BP82},
\begin{equation}
\dot m(r)=\dot m_{\rm in}\left[1+\left(\frac{\dot m_{\rm out}}{\dot m_{\rm in}}-1\right)\frac{\ln (r/r_{\rm in})}{\ln (r_{\rm out}/r_{\rm in})}\right],
\label{wind}
\end{equation}
where we have defined $\dot m\equiv \dot M c^2/L_{\rm Edd}$, $L_{\rm Edd}=4\pi G m_{\rm p}c/\sigma_{\rm T}$ is the Eddington luminosity for pure hydrogen, $m_{\rm p}$ is the proton mass and $\sigma_{\rm T}$ is the Thomson cross section (note that we do not include an efficiency in the definition of $\dot m$).
We use a constant $r_{\rm out}$, while $\dot m_{\rm out}/\dot m_{\rm in}\geq 1,\, \dot m_{\rm in},\, r_{\rm in}$ are free parameters. This prescription can give a relatively moderate increase of $\dot m$ with the increasing radius. For the case of no wind effect on $\dot m$ (which is then constant), we set $\dot m_{\rm out}/\dot m_{\rm in}= 1$.

\begin{figure}
\centerline{\includegraphics[width=7.5cm]{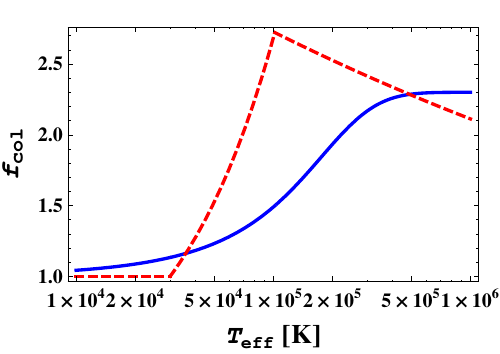}} 
\caption{The dependencies of $f_{\rm col}$ on the disk effective temperature using the formula of \chiang (blue solid curve) and that of \done (red dashed curve).}
\label{f:fcol}
\end{figure}

For the color correction, we use two options. In one, we use the formula of \citet{Chiang02} (hereafter \chiang), who fitted the numerical results given in fig.\ 13 of of \citet{Hubeny01},
\begin{equation}
f_{\rm col}(T_{\rm eff})=f_\infty-{(f_\infty-1)\left[1+\exp(-\nu_{\rm b}/\Delta\nu)\right]\over 1+\exp\left[(\nu_{\rm p}-\nu_{\rm b})/\Delta\nu\right]},
\label{fcol}
\end{equation}
where $\nu_{\rm p}=2.82 k T_{\rm eff}/h$, $f_\infty=2.3$, $\nu_{\rm b}=\Delta\nu=5\times 10^{15}$ Hz (in the source frame). This dependence is shown by the solid curve in Figure \ref{f:fcol}. We note it roughly agrees with the general trend shown in fig.\ 1 of \citet{Hubeny01}, while a constant value of $f_{\rm col}$ disagrees with the results of \citet{Hubeny01}. On the other hand, an $f_{\rm col}$ substantially greater than 1 is ruled out for emission at frequencies $\sim 10^{15}$ Hz in the source frame in the framework of the standard disk model, since it would predict a strong decrease of the flux (see equation \ref{nufnu}), not seen in the results of \citet{Hubeny01}. 

In the other option, we use the color-correction of \citet{Done12}, hereafter \done. It is given by
\begin{equation}
f_{\rm col}(T_{\rm eff})= \begin{cases}1, &T_{\rm eff}< 3\times 10^4{\rm K};\cr
\left(T_{\rm eff}\over 3\times 10^4{\rm K}\right)^{0.82}, &3\times 10^4{\rm K}\leq T_{\rm eff}\lesssim 10^5\,{\rm K};\cr
\left(72\,{\rm keV}\over k T_{\rm eff}\right)^{1/9}, &T_{\rm eff}\gtrsim 10^5{\rm K},\cr \end{cases}
\label{fcolD}
\end{equation}
which is shown by the dashed curve in Figure \ref{f:fcol}. (We note that equation 2 of \done is given in terms of the maximum disk temperature, $T_{\rm max}$, which should be instead the local effective temperature; C. Done, private communication.) The model using this correction is implemented as {\tt optxagnf}\footnote{\url{https://heasarc.gsfc.nasa.gov/xanadu/xspec/manual/node206.html}} in the {\sc xspec} \citep{Arnaud96} suite of spectral fitting routines, and we use it below. This model assumes a standard accretion disk including general-relativistic effects \citep{NT73}. Spectra from {\tt optxagnf} were tested against the models of \citet{Davis06}.

Then, a formula for the global hardening factors (as a function of $M$, $\dot M$ and the viscosity parameter) was given by equation (10) of \citet{Davis19}. They also presented some results as a function of the local $T_{\rm eff}$, but only for $T_{\rm eff}\gtrsim 2\times 10^5$\,K.

Since the flux can be measured at a different frequency, $\nu_1$, than that, $\nu_2$, of the size measurement (in particular from microlensing), we first solve equations (\ref{nufnu}--\ref{teff}) at $\nu_1$ for $\dot m_{\rm in}$ given $r_{\rm in}$ and $r_{\rm out}$, by setting $F_{\nu_1}$ equal to that observed. Then, the half-light radius at $\nu_2$ is given by $r_{1/2,\nu_2}$, which splits the integral in equation (\ref{nufnu}) into two halves (analogously to equation \ref{xhalf}). (In the case of the spectra of \done, we accessed the source code of {\tt optxagnf} and calculated $r_{1/2,\nu_2}$ based on that.) Our formalism, based on equations (\ref{nufnu}--\ref{teff}), yields only one disk size, and thus there is no more an ambiguity between the 'theory size' and the 'disk size'. 

\subsection{An example application}
\label{appl}

\begin{table}
\caption{The obtained values of $\dot m_{\rm in}$, $r_{1/2,\nu_2}$ and $L_{\rm disk}/L_{\rm Edd}$ for our models of SDSS 0924+0219}
\label{rhalf}      
\centering            
\begin{tabular}{cccccccc}
\hline 
\# & $r_{\rm in}$ & $r_{\rm b}$ & $f_{\rm col}$ & $\dot m_{\rm out}/\dot m_{\rm in}$ & $\dot m_{\rm in}$ & $r_{1/2,\nu_2}$ & $L_{\rm disk}/L_{\rm Edd}$\\
\hline
1 & 0  & 0 & 1       & 1 & 1.9 & 65  & --   \\ 
2 & 6  & 6 & 1       & 1 & 2.7 & 80  & 0.22 \\ 
3 & 6  & 6 & \chiang & 1 & 3.5 & 107 & 0.28 \\
4 & 6  & 6 & \done   & 1 & 3.2 & 107 & 0.21 \\
5 & 20 & 6 & \chiang & 1 & 3.6 & 113 & 0.17 \\
6 & 20 & 6 & \done   & 1 & 3.2 & 109 & 0.14 \\
7 & 6  & 6 & \chiang & 5 & 0.9 & 116 & 0.17 \\
8 & 20 & 6 & \chiang & 5 & 1.1 & 128 & 0.10 \\
\hline                    
\end{tabular} 
\end{table}

As an example illustrating the above formalism, we use the parameters of the quasar SDSS 0924+0219 ($z=1.524$) as given by \citet{Morgan10}\footnote{We note that \citet{Morgan10} gave their estimates of the radii from microlensing, $R_{\rm S}$ in their notation, as $R_\nu$, not $R_{1/2,\nu}$.}, namely $M=1.1\times 10^8\msun$, $D_L\approx 11.1$\,Gpc, $\nu_1\approx 3.7\times 10^{14}$\,Hz (the middle of the $I$ band), $F_{\nu_1}\approx 7.7\times 10^{-29}$\,erg\,cm$^{-2}$\,s$^{-1}$\,Hz$^{-1}$, $\nu_2\approx 4.8\times 10^{14}$\,Hz [$\lambda_2=2500(1+z)$\,\AA]. The half-light radius at $\nu_2$ from microlensing following from the estimate by \citet{Morgan10} is $r_{\rm 1/2,obs}\approx 150$. 

\begin{figure}
\centerline{\includegraphics[width=8cm]{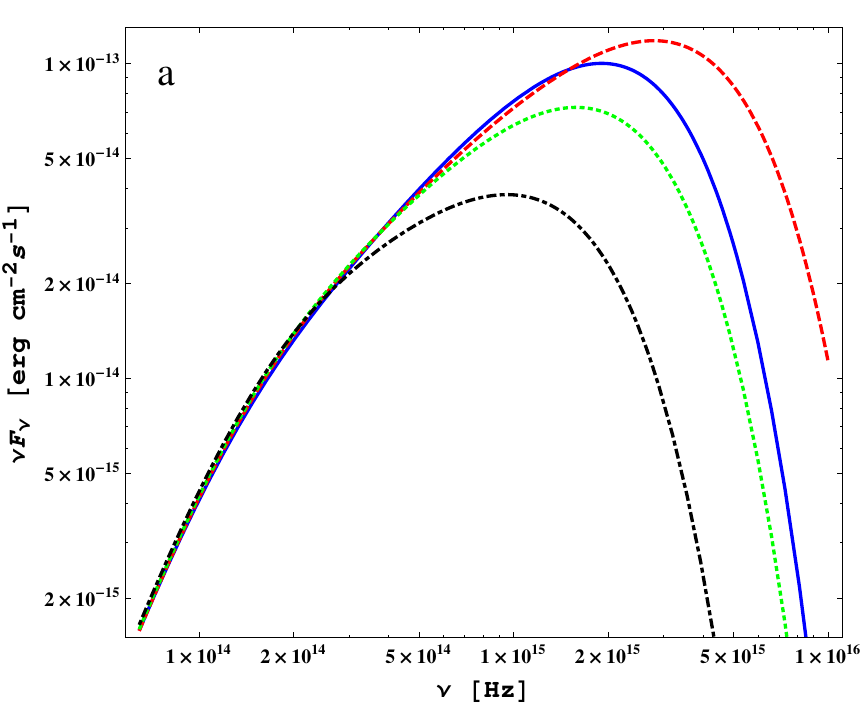}} 
\centerline{\includegraphics[height=8cm, angle=-90]{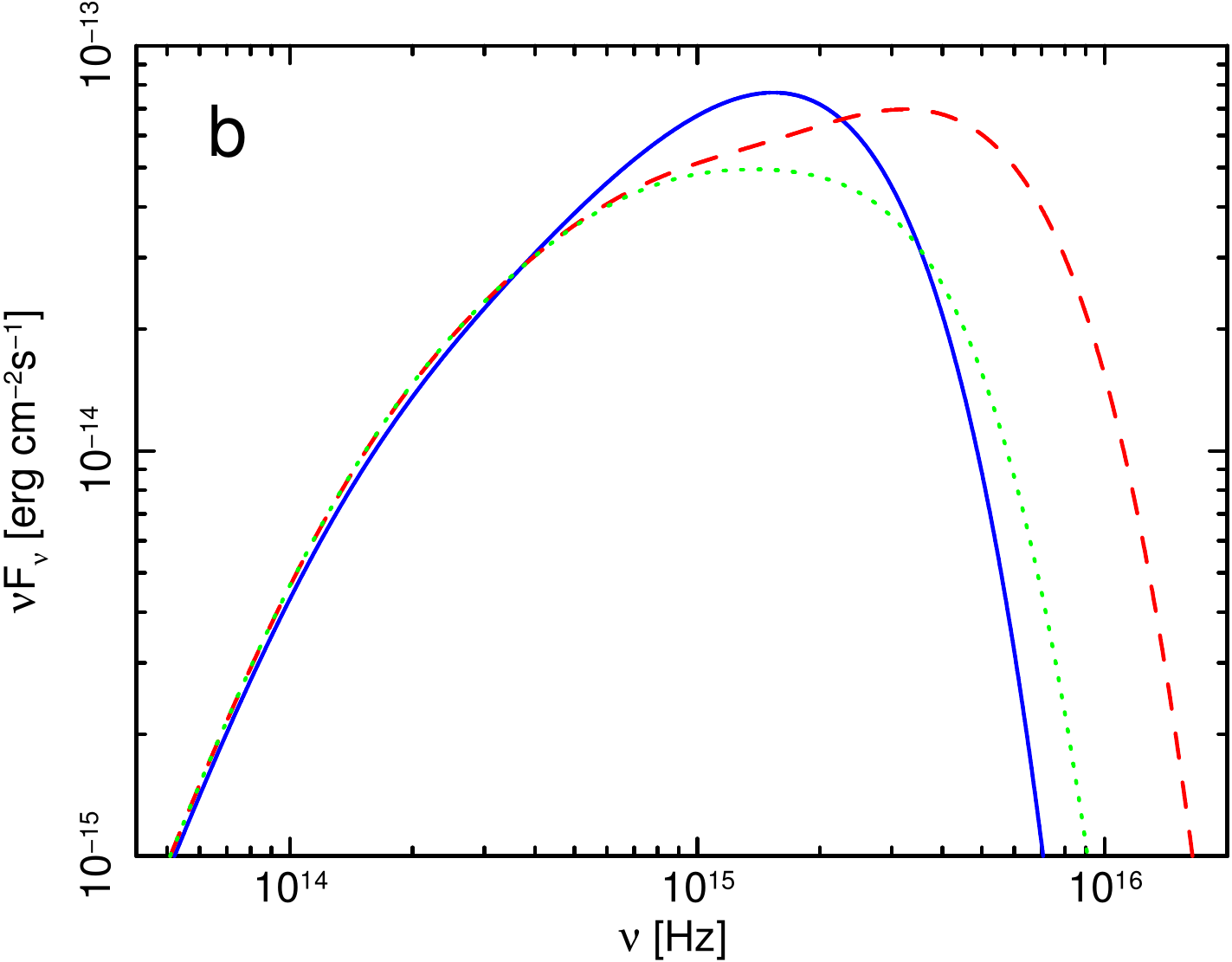}} 
\caption{The disk blackbody spectra shown at the observer frame for $r_{\rm b}=6$ for the parameters of SDSS 0924+0219 \citep{Morgan10} for some models of Table \ref{rhalf}. (a) The blue solid and red dashed curves correspond to $r_{\rm in}=6$ and $f_{\rm col}=1$ and $f_{\rm col}$ of \chiang, respectively. The green dotted and black dot-dashed curves give the spectra for $r_{\rm in}=20$, $f_{\rm col}$ of \chiang and $\dot m_{\rm out}/\dot m_{\rm in}=1$ and 5, respectively. (b) The disk blackbody spectra obtained with the models of \done ($\dot m_{\rm out}/\dot m_{\rm in}=1$). The blue solid curve is for $f_{\rm col}=1$, $r_{\rm in}=6$, which was obtained with the {\tt optxagn} model. The red dashed and green dotted curves are for $f_{\rm col}$ of {\tt optxagnf} (\done) and $r_{\rm in}=6$ and 20, respectively. All of the spectra are normalized to the observed flux at $\nu_1=3.7\times 10^{14}$\,Hz, where all of the above spectra intersect.
}
\label{spectra}
\end{figure}

\begin{figure}
\centerline{\includegraphics[width=8cm]{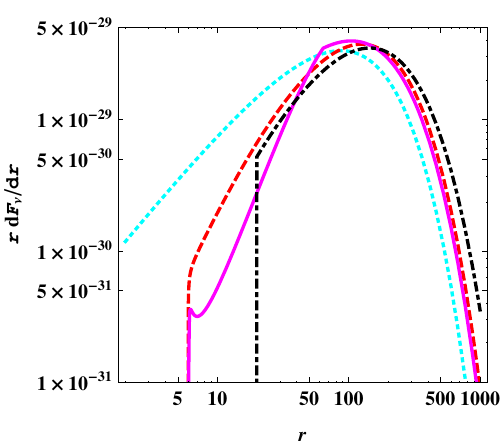}} 
\caption{The radial emissivity profiles, ${\rm d} F_\nu /{\rm d}\ln r$, at the source-frame wavelength of 2500\,\AA\ for some models of Table \ref{rhalf}. The dotted cyan curve is for $r_{\rm in}=0$, $\dot m_{\rm out}/\dot m_{\rm in}=1$, and $f_{\rm col}=1$. The red dashed and magenta solid curves are for $r_{\rm in}=6$ and the color corrections of  \chiang and \done, respectively. The black dot-dashed curve is for $r_{\rm in}=20$, $\dot m_{\rm out}/\dot m_{\rm in}=5$ and the color corrections of \chiang. }
\label{profiles}
\end{figure}

We consider a sequence of models with diferent assumptions about the inner radii, the color correction, and $\dot m_{\rm out}/\dot m_{\rm in}$. All of the models are normalized to yield the observed flux at $\nu_1$, and in all of them we assume $r_{\rm out}=10^3$. In models with an inner boundary condition, we assume the dimensionless spin parameter of 0. Following \citet{Morgan10}, we assume $\cos i=0.5$. Table \ref{rhalf} lists the model parameters including the total disk luminosity, Figure \ref{spectra} shows some of the obtained spectra, and Figure \ref{profiles} shows selected radial profiles of the emissivity at the frequency at which the microlensing size was measured, $\nu_2$. 

We begin with model 1 with $f_{\rm col} =1$ and $r_{\rm in}=r_{\rm b}=0$, corresponding to the neglect of the inner boundary condition (as in Equations \ref{rnu2}--\ref{xhalf}). We see in Table \ref{rhalf} that its half-light radius, $\approx$65, is much lower than $r_{\rm 1/2,obs}$. This is reflected in the emissivity profile, shown in Figure \ref{profiles}, which has a strong low-radius tail. Our model 2 is similar except that it includes the inner boundary term, with $r_{\rm in}=6$. This increases $r_{1/2,\nu_2}$ to $\approx$80. Then our models 3 and 4 show the effect of using the color corrections of \chiang and \done, respectively. In spite of relatively different spectra of those models, see Figure \ref{spectra}, $r_{1/2,\nu_2}\approx 107$ in both cases. 

We then examine the effect of a disk truncation. As found, e.g., by \citet{Mosquera13}, the sizes of the X-ray sources in quasars estimated by microlensing are $\sim 20 R_{\rm g}$. Thus, it is reasonable to assume $r_{\rm in}=20$ as a disk truncation radius, which we assume in our models 5 and 6. However, this only slightly increases $r_{1/2,\nu}$. Finally, models 7 and 8 include a disk wind, with $\dot m_{\rm out}/\dot m_{\rm in}=5$. Model 8 also includes the disk trunction, where we obtain $r_{1/2,\nu_2} \approx 128$, which is relatively close to the observed value, and it can approach it for a larger $\dot m_{\rm out}/\dot m_{\rm in}$. We note that the accretion rate supplied to the outer edge of the disk is given by $(\dot m_{\rm out}/\dot m_{\rm in})\dot m_{\rm in}\approx 4.5$--5.5, which is then only slightly larger than the values of $\dot m_{\rm out}=\dot m_{\rm in}\approx 3.2$--2.6 for the models with color correction and without disk winds. 

At low frequencies in Figure \ref{spectra}, we see that using the outer disk radius of $r_{\rm out}=10^3$ hardens the spectral slope. Thus, there is not any extended range with $F_\nu \propto\nu^{1/3}$, assumed in the approximations of equations (\ref{rnu1}) and (\ref{rnu2}). In Figure \ref{profiles}, we see that using the color corrections substantially enhances the relative contributions of outer disk regions. 

Thus, we have shown that the combined effects of the color correction, disk truncation and moderate disk wind can lead to a relatively good agreement of the theoretical disk radii with those estimated by microlensing.

\subsection{Additional effects}
\label{add}

There are two additional effects which could further increase the theoretical sizes, and have not been considered in this work. One is that the disk may be covered by an X-ray emitting corona, in which a fraction of the gravitational energy is dissipated as well as it upscatters a fraction of the disk photons to higher energies, thus reducing the direct disk flux \citep{SZ94, Gierlinski99, Done06,You16b}. If a fraction, $f_{\rm c}<1$, of the disk emission is lost due to both covering by a corona and the coronal dissipation, a multiplicative factor of $(1-f_{\rm c})$ will appear in Equation (\ref{nufnu}), and the estimated disk size will be increased by $(1-f_{\rm c})^{-1/2}$. This coefficient can be defined to include the effect of disk dissipation reduction being partially offset by the disk irradiation by the corona and the subsequent quasi-thermal re-emission. 

A second related effect is the irradiation of outer disk regions by the central X-ray source (e.g., \citealt{Kammoun21b, Kammoun21a}). This also leads to an increase of the estimated disk sizes. However, the X-ray fluxes in microlensing quasars are relatively low, and we expect this effect to be minor.

Finally, we note that \citet{Dai10} made a comparison of the disk sizes at a given frequency (both $R_\nu$ and $R_{1/2,\nu}$) between the standard disk model \citep{SS73} and the model of \citet{Hubeny01}, either relativistic and non-relativistic, and either assuming the local blackbody and with the NLTE disk atmosphere. They found that the model of \citet{Hubeny01} gives generally {\it smaller\/} radii at an assumed value of $\dot M$ than equation (\ref{rnu1}). However, it appears that \citet{Dai10} defined $R_\nu$ by $h\nu=k T_{\rm eff}$, without including the color correction. In general, the blackbody emission is the most efficient radiation process possible (in the absence of coherent processes), and any departure from the full thermodynamical equilibrium has to lead to an increase of the emitting area.

\section{Conclusions}
\label{conclusions}

We have studied the effects of including a color correction factor (based on the results of \citealt{Hubeny01}, \citealt{Davis06} and \done), a disk truncation at an inner radius $>$ISCO, and of a decrease of $\dot M$ with the decreasing disk radius due to winds on the half-light radii of standard accretion disks. In the case of the color correction, we have found it is important to estimate the size using half-light radii rather than the radii based on monochromatic approximations, equations (\ref{rnu1}) and (\ref{rnu2}). The cause of this is that while $f_{\rm col}\approx 1$ at $R_\nu$, it increases fast at lower radii, which, in turn, moves the emissivity radial profile to larger radii, see Figure \ref{profiles}. In the considered example, it increases the half-light radius from $65 R_{\rm g}$ calculated with the widely-used simplified formulae to about $\sim 110 R_{\rm g}$. Then, including the effect of an inner truncation radius of the disk being above the ISCO (compatible with the presence of X-ray emission) leads to a further increase. Thus, these two effects are important, and should be included in interpreting the microlensing results.

Furthermore, we have considered the effect of disk winds, which can cause the local accretion rate to decrease with the decreasing radius. When combined with the effect of truncation, we find the half-light radius to increase to about $130 R_{\rm g}$ for a modest $\dot M_{\rm out}/\dot M_{\rm in}=5$. This is similar to the half-light radius from microlensing, of $\approx\! 150 R_{\rm g}$ in the considered example of SDSS 0924+0219 \citep{Morgan10}.

The formalism used in this paper gives also the mass accretion rate of a given model, removing the ambiguity between the 'theory' and 'flux' radii. For the model with the disk truncation and a constant accretion rate, the implied value is $\dot m_{\rm in}\approx 0.3$--0.4. For an assumed accretion efficiency of 0.1, this corresponds to about 0.3--0.4 of the Eddington luminosity, in agreement with the estimates of the Eddington ratios of quasars of \citet{Kollmeier06}.

Finally, we note that the above results were derived assuming the standard optically-thick accretion disk, i.e., one in which the energy dissipation occurs close to the disk midplane. As we mentioned in Section \ref{intro}, accretion disks supported by magnetic pressure are substantially hotter than the standard ones, and thus have substantially larger characteristic sizes at a given emitted frequency (e.g., \citealt{Begelman07,Begelman17}), which provides an alternative solution to the problem of the accretion disk sizes.

\section*{Acknowledgments}
We thank E. Agol, E. Kammoun, Ch.\ Kochanek, Ch.\ Morgan, and F. Yuan for valuable discussions, and the referee of this paper, Chris Done, for important suggestions and comments. We acknowledge support from the Polish National Science Center under the grants 2019/35/B/ST9/03944 and 2016/21/B/ST9/02388, and from the Natural Science Foundation of China (360 U1931203, 11903024 and 12273026) and the National Key Research and Development Program of China (2021YFA0718500). Our work also benefitted from discussions during Team Meetings in the International Space Science Institute (Bern).

\appendix
\section{Simplified formulae}
\label{simple}

We can define the radius, $R_\nu$, at which the local color temperature corresponds to the source-frame frequency, $h\nu(1+z)= k T_{\rm eff}(R_\nu) f_{\rm col}\left[T_{\rm eff}(R_\nu)\right]$. If we are concerned with emission at $R_\nu\gg R_{\rm in}$, the second factor in equation (\ref{teff}) can be neglected, in which case $T_{\rm eff}\propto R^{-3/4}$. The expression for $R_\nu$ in terms of $M$ and $\dot M$ can be then obtained
\begin{equation}\label{rnu1}
R_{\nu,T} \approx \left[k f_{\rm col}(R_{\nu,1})\over h \nu(1+z)\right]^{4/3}
\left[ 3 G M \dot M(R_{\nu,1}) \over 8 \pi \sigma\right]^{1/3}.
\end{equation}
We note that this expression, often called the 'theory size', depends on $\dot M(R_\nu)$, which is usually not directly known.

If $\dot M$ and the color correction are constant, we have $k T(x)= h \nu(1+z) x^{-3/4}$, where $x\equiv R/R_\nu$. We can then rewrite equation (\ref{nufnu}) as
\begin{align}
&\nu F_\nu \approx   {4 \pi h\nu^4 (1+z)^4 R_\nu^2 I_1 \cos i \over {f_{\rm col}^4 D_L^2 c^2}},\label{fnu2}\\ 
&I_1\equiv \int^\infty_0\!\!\! {x \,{\rm d}x\over \exp(x^{3/4})-1}=\frac{4}{3}\Gamma\left(\frac{8}{3}\right) \zeta\left(\frac{8}{3}\right) \approx 2.5762,
\end{align}
where $\Gamma$ and $\zeta$ are the Euler Gamma function and the Riemann zeta function, respectively. This yields
\begin{equation}
R_{\nu,F} \approx   {f_{\rm col}^2 D_L c (\nu F_\nu)^{1/2}\over 2 \nu^2 (1+z)^2(\pi I_1 h \cos i)^{1/2} }, \label{rnu2}
\end{equation}
which is often called the 'flux size'. Note that in the considered limit, $R_{\nu,F}$ is independent of the mass, depends only on $F_\nu$ and $i$, and is $\propto f_{\rm col}^2$. We also note that the ratio $R_{\nu,F}/R_{\nu,T}\propto f_{\rm col}^{2/3}$; thus it is larger than unity for $f_{\rm col}>1$, which can explain values of $R_{\nu,T}$ being larger than $R_{\nu,F}$ at the assumed $f_{\rm col}=1$ in microlensed quasars (e.g., \citealt{Morgan10}). If we equate $R_{\nu,T}=R_{\nu,F}$, we obtain the expression for the color-corrected disk blackbody spectrum in the intermediate region, $F_\nu\propto \nu^{1/3}$, between the low and high energy cutoffs due to the disk outer and inner radii, respectively,
\begin{equation}
\nu F_\nu= \frac{ \pi^{1/3}I_1 k^{8/3} \cos i\, (3  G M \dot{M}/\sigma)^{2/3}}{c^2 D_L^2  h^{5/3}}\left[\frac{(1+z) \nu}{f_{\rm col}}\right]^{4/3}.
\label{diskbb}
\end{equation}
A discrepancy between the values of $R_{\nu,T}$ and $R_{\nu,F}$ may indicate that either the local emission is not given by the disk blackbody spectrum, equation (\ref{diskbb}), or $\dot M$ and/or $f_{\rm col}$ are not properly estimated.

For constant $\dot M$ and $f_{\rm col}$, the half-light radius corresponds to the solution of 
\begin{equation}
\int^{x_{1/2}}_0\!\!\!\!\! {x \,{\rm d}x\over \exp(x^{3/4})-1}=
\int^\infty_{x_{1/2}}\!\! {x \,{\rm d}x\over \exp(x^{3/4})-1},\,\, x_{1/2}\approx 2.4356,
\label{xhalf}
\end{equation}
i.e., $R_{1/2,\nu}={x_{1/2}} R_{\nu,F}$. In the considered limit, $R_{1/2,\nu'} =(\nu'/\nu)^{-4/3} R_{1/2,\nu}$. 

\bibliographystyle{../aasjournal}
\bibliography{../allbib}

\end{document}